\documentclass[%
 reprint,
superscriptaddress,
nofootinbib,
 amsmath,amssymb,
 aps,
prl,
floatfix,
]{revtex4-2}
\usepackage[hidelinks]{hyperref}

\usepackage[T1]{fontenc} % if needed
\usepackage{tikz}
\usepackage{pifont}% http ://ctan.org/pkg/pifont
\usepackage{amsmath} % for \text
\usepackage{mathrsfs} % for \mathscr
\usepackage[scr=boondox]{mathalpha}
\usepackage{xfp} % higher precision (16 digits?)
\usepackage[outline]{contour} % glow around text
\usetikzlibrary{decorations.markings,decorations.pathmorphing}
\usetikzlibrary{angles,quotes} % for pic (angle labels)
\usetikzlibrary{arrows.meta} % for arrow size
\usetikzlibrary{external}
\contourlength{1.4pt}

\usepackage{braket}
\usepackage{bbm}

\usepackage{amsmath, amssymb, amsfonts, amsthm, latexsym, epsfig, mathrsfs, xcolor, slashed, braket, cancel}
\usepackage[overload]{textcase}

\usepackage{cleveref}
\usepackage{soul}
\usepackage{yfonts}

\newcommand{\Tr}{\textrm{Tr}}
\tikzset{>=latex} % for LaTeX arrow head
\colorlet{myred}{red!80!black}
\colorlet{myblue}{blue!80!black}
\colorlet{mygreen}{green!80!black}
\colorlet{mydarkred}{red!50!black}
\colorlet{mydarkblue}{blue!50!black}
\colorlet{mylightblue}{mydarkblue!6}
\colorlet{myvlightblue}{mydarkblue!3}
\colorlet{mypurple}{blue!40!red!80!black}
\colorlet{mydarkpurple}{blue!40!red!50!black}
\colorlet{mylightpurple}{mydarkpurple!80!red!6}
\colorlet{myorange}{orange!40!yellow!95!black}
\tikzstyle{cone}=[mydarkblue,line width=0.2,top color=blue!60!black!30,
 bottom color=blue!60!black!50!red!30,shading angle=60,fill opacity=0.9]
\tikzstyle{cone back}=[mydarkblue,line width=0.1,dash pattern=on 1pt off 1pt]
\tikzstyle{world line}=[myblue!60,line width=0.4]
\tikzstyle{world line t}=[mypurple!60,line width=0.4]
\tikzstyle{particle}=[mygreen,line width=0.5]
\tikzstyle{photon}=[-{Latex[length=4,width=3]},myorange,line width=0.4,decorate,
 decoration={snake,amplitude=0.9,segment length=4,post length=3.8}]
\tikzstyle{singularity}=[myred,line width=0.6,decorate,
 decoration={zigzag,amplitude=2,segment length=6.17}]
\tikzset{declare function={%
 penrose(\x,\c) = {\fpeval{2/pi*atan( (sqrt((1+tan(\x)^2)^2+4*\c*\c*tan(\x)^2)-1-tan(\x)^2) /(2*\c*tan(\x)^2) )}};%
 penroseu(\x,\t) = {\fpeval{atan(\x+\t)/pi+atan(\x-\t)/pi}};%
 penrosev(\x,\t) = {\fpeval{atan(\x+\t)/pi-atan(\x-\t)/pi}};%
 kruskal(\x,\c) = {\fpeval{asin( \c*sin(2*\x) )*2/pi}};% Penrose coordinates for Kruskal
}}

\crefname{section}{sec.}{sec.}
\crefname{appendix}{Appendix}{Appendices}
\crefname{table}{Table}{Tables}

\crefname{definition}{Def.}{Defs.}
\crefname{prop}{Prop.}{Props.}
\crefname{lemma}{Lemma}{Lemmas}
\crefname{corollary}{Cor.}{Cors.}
\crefname{thm}{Theorem}{Theorems}
\crefname{remark}{Remark}{Remarks}

\crefname{ass}{Assumptions}{Assumptions}
\crefname{property}{Properties}{Properties}

\newcommand{\be}{\begin{equation}\begin{aligned}}
\newcommand{\ee}{\end{aligned}\end{equation}}

\newcommand{\mc}{\mathcal}

\newcommand{\ms}{\mathscr}
\newcommand{\mf}{\mathfrak}
\newcommand{\bb}{\mathbb}

\newcommand{\normord}[1]{:\mathrel{#1}:}

\usepackage{enumitem}

\setlist[enumerate]{itemsep=2pt, label=(\arabic*), ref=(\arabic*)}

\definecolor{indigo(dye)}{rgb}{0.0, 0.25, 0.42}

\newcommand{\defn}{\mathrel{\mathop:}=} %shrtct for definition operator

\newif\ifslow
\newcommand{\op}[1]{\boldsymbol{#1}}

\newcommand{\Alg}{\mathscr{A}}

\newcommand{\s}{\omega}

\newcommand{\Hilb}{\mathscr{H}}

\newcommand{\antiHilb}{%
\hspace{4pt} % puts a leading space which Latex skips with below code
 \vbox{%
 \hrule height 0.5pt% % Line above with certain width
 \kern0.25ex% % Distance between line and content
 \hbox{%
 \kern-0.3em% % Distance between content and left side of box, negative values for lines shorter than content
 \ifmmode\Hilb\else\ensuremath{\Hilb}\fi% % The content, typeset in dependence of mode
 \kern0em% % Distance between content and right side of box, negative values for lines shorter than content
 }% end of hbox
 }% end of vbox
}

\newcommand{\Fock}{\mathscr{F}}

\let\oldsetminus\setminus
\renewcommand{\setminus}{\!\oldsetminus\!} 

\let\oldint\int
\renewcommand{\int}{\oldint\limits}

\let\oldlim\lim
\renewcommand{\lim}{\oldlim\limits}

\renewcommand{\bar}{\overline}

\newcommand\sH{{\ensuremath{{\mathcal H}}}}

\begin{document}

\title{
Algebraic Observational Cosmology}

% \preprint{MIT-CTP/5332}

\author{Jonah Kudler-Flam}
\affiliation{School of Natural Sciences, Institute for Advanced Study, Princeton, NJ 08540, USA}
\affiliation{Princeton Center for Theoretical Science, Princeton University, Princeton, NJ 08544, USA}
\author{Samuel Leutheusser}
\affiliation{Princeton Gravity Initiative, Princeton University, Princeton NJ 08544, USA} 
\author{Gautam Satishchandran}
\affiliation{Princeton Gravity Initiative, Princeton University, Princeton NJ 08544, USA}   

\begin{abstract}
What can be measured by an observer in our universe? We address this question by constructing an algebra of gravitationally-dressed observables accessible to a comoving observer in FLRW spacetimes that are asymptotically de Sitter in the past, describing an inflationary epoch. 
An essential quantized degree of freedom is the zero-mode of the inflaton, which leads to fluctuations in the effective cosmological constant during inflation and prevents the existence of a maximum entropy state in the semiclassical limit. Due to the inaccessibility of measurements beyond our cosmological horizon, we demonstrate that all states are mixed with well-defined von Neumann entropy (up to a state-independent constant).
For semiclassical states, the von Neumann entropy corresponds to the generalized entropy of the observer's causal diamond, a fine-grained quantity that is sensitive to the initial conditions of the universe.
\noindent 
 
\end{abstract}

\date{\today}
\maketitle

\paragraph*{Introduction.---}
Our universe is, to a very good approximation, well-described by a spatially homogeneous and isotropic spacetime \cite{Green:2014aga}. Indeed, the $\Lambda$-CDM model of cosmology successfully accounts for structure
formation \cite{2002PhR...367....1B}, the abundance
of light elements \cite{2016RvMP...88a5004C}, cosmic acceleration \cite{perlmutter1999measurements, riess1998observational}, as well as the statistical properties of the cosmic microwave background seeded from inflation \cite{baumann2022cosmology,2020A&A...641A...6P}. The cosmological ansatz of isotropy and homogeneity, leads to the FLRW spacetimes with metric
\begin{align}
\label{eq:FLRW}
    ds^2 = -dt^2 + a(t)^2\left( dr^2 + r^2 d\Omega^2\right),
\end{align}
where $a(t)$ is the dimensionless scale factor.

\begin{figure}
    \centering
    \includegraphics[]{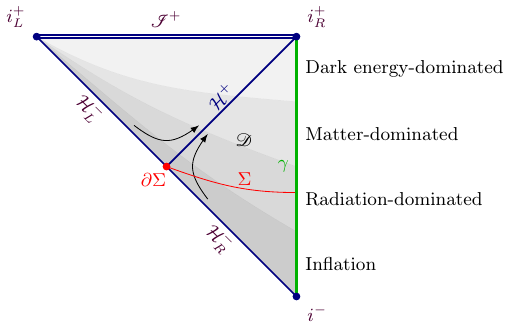}
    \caption{The Penrose diagram for FLRW spacetimes that model our universe seeded from inflation. 
    $\ms{D}$ is the domain of dependence for the observer $\gamma$, $\mc{H}^{+}$ is the future horizon of $\gamma$, $\mc{H}^{-}_{\textrm{R}}$ is the past horizon, and $\Sigma$ is a Cauchy surface for $\ms{D}$. 
    }
    \label{fig:penrose_tikz}
\end{figure}

We are cosmological observers, traveling approximately on a geodesic through a continuously evolving spacetime. In principle, we are able to measure all quantum field observables along our timelike worldline, $\gamma,$ (see Figure~\ref{fig:penrose_tikz}). For observers in Minkowski spacetime, the ability to make all measurements along an infinitely extended worldline implies the ability to measure the global wavefunction of the universe \cite{borchers1961vollstandigkeit, araki1963generalization,2024AnHP..tmp...14S}. However, the positivity of the cosmological constant in our universe implies that we experience a cosmological horizon, $\mc{H}^+$, beyond which we cannot, even in principle, make measurements. This
uncertainty in our knowledge of the universe can be quantified by an \textit{entropy}. It is therefore fundamental to ask what is the class of observables that we can measure along our worldline and what is the associated entropy? In this Letter, we address this question in the context of semiclassical quantum gravity by constructing an explicit algebra of gravitationally-dressed observables localized to the observable universe, $\mathscr{D}$, and demonstrate that the entropy is given by a finite ``generalized entropy,'' consisting of the sum of the area of the cosmological horizon (in Planck units with $\hbar = k_B= c = 1$) during inflation and the von Neumann entropy of quantum fields
\begin{align}
\label{eq:sgen}
    S_{\text{observer}} = \frac{\text{Area}(\partial \Sigma)}{4G_{\textrm{N}}} + S_{\textrm{vN}}(\Sigma).
\end{align}
This formula is consistent with general expectations regarding entropy in quantum gravity \cite{2002RvMP...74..825B}, though has yet to be derived from first principles for cosmological spacetimes. 

We take a bottom-up approach, quantizing the gravitational fluctuations, matter fields, and {a simple model of} the observer in perturbation theory. In quantum gravity, there are nontrivial constraints between the metric and matter fields due to Einstein's equations.
By imposing these constraints, we obtain an algebra of gravitationally-dressed observables in $\ms{D}$ with a novel structure. Unlike non-gravitational quantum field theory algebras where traces do not exist and von Neumann entropies diverge (see e.g.~\cite{2018arXiv180304993W} for a review), the 
algebra of gravitationally-dressed observables that we 
obtain has a well-defined trace and associated finite von Neumann entropies. Our construction generalizes recent developments on {the} algebra {of observables} that emerge in semiclassical quantum gravity \cite{2022JHEP...10..008W,Chandrasekaran:2022cip,2022arXiv220910454C,Jensen:2023yxy,2023arXiv230915897K,2023arXiv230803663W}. In particular, our analysis clarifies the algebraic structure of quantum {gravitational} observables in general backgrounds.  The novel ingredient 
is the quantization of a ``zero-mode'' that labels a continuous family of solutions to Einstein's equations  (e.g., the amplitude of a gravitating electromagnetic wave \cite{2023arXiv230803663W} or the mass of a black hole \cite{2022arXiv220910454C,2023arXiv230915897K}). In cosmological spacetimes, we find that this zero-mode is straightforwardly understood as fluctuations in the value of the inflaton field at the beginning of inflation, which is directly related to fluctuations in the effective cosmological constant.

\paragraph*{Set Up and Classical Background.---}
To give a complete description of the universe accessible to the observer, we should consider all cosmologically relevant degrees of freedom including the standard model, the inflaton, and dark matter.  However, as we will see, the essential features of the algebra of observables can be obtained in a minimal cosmological model composed of only the metric $g_{ab}$, the inflaton $\phi$, and an observer with energy $\varepsilon$. Incorporating other degrees of freedom is straightforward and is detailed in the supplemental material (SM). Thus, the model we consider is governed by equations of motion
\begin{align}
\label{eq:EE}
    G_{ab} + \Lambda g_{ab} &= 8\pi G_{\textrm{N}} T_{ab} ,
    \\
\label{eq:KG}
    \Box_g \phi &= \partial_{\phi}V(\phi),
    \\
    u^{a}\nabla_{a}\tau&=-1  ,\quad u^{a}\nabla_{a}\varepsilon= 0. \label{eq:obs}
\end{align}
where $\tau$ is the proper time of the observer, $\Lambda$ is a cosmological constant, $V$ is the potential of the inflaton, {$u^{a}$ is the unit-normalized tangent vector to $\gamma$, $\nabla_{a}$ is the covariant derivative and $\Box_{g}=g^{ab}\nabla_{a}\nabla_{b}$. The first equation in \eqref{eq:obs} is manifestly reparametrization invariant and the second equation states that the energy is constant along the observer wordline.}
The stress-energy tensor $T_{ab}$ on the right-hand-side of \eqref{eq:EE} is given by the sum of the inflaton stress-energy 
\begin{equation}
T_{ab}^{(\phi)} = \nabla_{a}\phi \nabla_{b}\phi - g_{ab}\bigg(\frac{1}{2}\nabla_{c}\phi \nabla^{c}\phi + V(\phi)\bigg)
\end{equation}
and the observer stress-energy 
\begin{equation}
T_{ab}^{(\varepsilon)} = \varepsilon u_{a}u_{b}\delta_{\gamma}
\end{equation}
where $\delta_{\gamma}$ is a delta function along the worldline.

\begin{figure}
    \centering
    \includegraphics[]{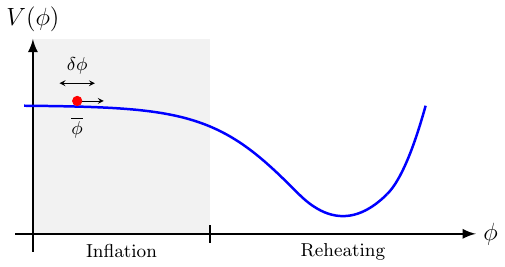}
    \caption{Sketch of an inflaton potential with slow roll region (shaded in grey).}
    \label{toy_tikz}
\end{figure}

{We now define our classical background, by solving \eqref{eq:EE}--\eqref{eq:obs} at leading order as  $G_{\textrm{N}} \to 0$.} This provides the background value of fields, whose fluctuations will subsequently be quantized. During inflation, the universe was approximately de Sitter, so we will restrict to background FLRW metrics $\bar{g}_{ab}$ that are {asymptotically} de Sitter in the far past (i.e.,  $a(t) \rightarrow e^{H_0 t}$ as $t\rightarrow -\infty$) where $H_0$ is the Hubble parameter.
Under the cosmological ansatz, the background inflaton field $\bar{\phi}$ only depends on time, so the background stress-energy is
\begin{align}
\label{eq:barT}
    \bar{T}_{ab}^{(\phi)} = (\partial_t \bar{\phi})^2t_{a}t_{b}-\bar{g}_{ab}\left(-\frac{1}{2} (\partial_t \bar{\phi})^2 + V(\bar{\phi})\right).
\end{align}
where $t_{a}=\nabla_{a}t$. We note that any slowly rolling solution $\bar{\phi}$ (i.e. $\bar{\phi}$ satisfying $(\partial_t \bar{\phi})^2 \ll V(\bar{\phi})$)
acts as an effective cosmological constant in Einstein's equations. 
For the stress-energy to have a non-trivial backreaction as $G_{\textrm{N}}\to 0$, we take the potential to behave as 
\begin{align}
\label{eq:V(phi)}
    V(\phi)\simeq \frac{\Lambda_{\textrm{eff}}}{8\pi G_{\textrm{N}}} + V'\phi + O(\sqrt{G_{\textrm{N}}}) \quad \textrm{  (early times)}
\end{align}
as $\phi \rightarrow -\infty$. In this regime, both slow-roll parameters, $\epsilon \sim (\partial_{\phi}V/V)^2/G_{\textrm{N}}$ and $\eta \sim \partial^2_{\phi}V/(G_{\textrm{N}}V)$, are suppressed by $G_\textrm{N}$. Here, $V^{\prime}$ is a constant parameter which determines the slope of the potential (see figure \ref{toy_tikz}).
This yields the early time Hubble parameter 
\begin{align}
    H_0 = \sqrt{\frac{\Lambda+\Lambda_{\textrm{eff}}}{3}}.
\end{align}
The observer energy is naturally $O(G_{\textrm{N}}^0)$ and so does not backreact in the semiclassical limit. Thus, the observer's worldline is a geodesic on $\bar{g}_{ab}$, $u^{a}=t^{a}$, and the observer proper time is the FLRW time (i.e. $\tau=t$).

The Klein-Gordon equation~\eqref{eq:KG} then becomes
\begin{align}
    \partial_t^2 \bar{\phi} + 3H_0 \partial_t\bar{\phi} = -\partial_{\phi}V(\bar{\phi})
\end{align}
which, indeed, admits slow-roll solutions\footnote{Whether $\bar{\phi}$ has a limit as $G_{\textrm{N}} t\to -\infty$ depends on the details of the potential at large negative values of $\bar{\phi}$. The future evolution of the background is also potential dependent, with some choices leading to reheating occurring in $O(G_{\textrm{N}}^0)$ time, while others result in a time that inversely scales with $G_{\textrm{N}}$.
} 
\begin{align}
\label{eq:phbar}
    \bar{\phi}(t) \simeq -\frac{V'}{3H_0}t + \bar{\phi}(0) \quad \textrm{ { (early times).}}
\end{align}
We note that our analysis is unaffected by the complicated details of reheating \cite{2006RvMP...78..537B} and our results depend only on the early time behavior of $\bar{\phi}$ given by \eqref{eq:phbar}. 

\paragraph*{Quantum Fluctuations and the Zero-Mode.---}

{With a classical background in hand, we can now analyze quantum fluctuations. We will focus on the quantization of scalar fluctuations in spatially-flat gauge. The quantization of (decoupled) linearized gravitons can be straightfowardly included and is treated in the SM to avoid repetition. Their inclusion --- along with any other matter fields --- does not affect our main points in any significant way.}

We consider small fluctuations{, $\delta \phi,$} about the spatially homogeneous classical solution {$\bar{\phi},$ i.e.} \cite{1992PhR...215..203M}
\begin{align}
    \phi(x,t) = \bar{\phi}(t)+ \delta\phi (x,t),
\end{align}
where we work in spatially flat gauge.
The sense in which these are ``small'' is that they are quantum fluctuations with typical size $O(\hbar^{1/2}H_0)$ in one Hubble time \cite{vilenkin1982gravitational,linde1982scalar,starobinsky1982dynamics}, which is much smaller than the field distance traveled by the background of $O(V'/H_0^2)$. 
This is directly related to setting conditions such that eternal inflation does not occur \cite{2007JPhA...40.6811G}. We can thus treat $\delta$ formally as an $\hbar^{1/2}$ expansion.
The perturbation obeys 
\begin{align}
\label{eq:deltaphi}
    \Box_{{\bar{g}}} \delta\phi(x,t)= V^{\prime \prime} (\bar{\phi})\delta\phi(x,t)
\end{align}
where $\Box_{{\bar{g}}}$ is the wave-operator on $\bar{g}_{ab}$. By \eqref{eq:V(phi)}, $V^{\prime \prime}\to 0$ as $t\to -\infty$ and so $\delta \phi$ satisfies a massless, minimally coupled wave equation at asymptotically early times.

We now quantize $\delta \phi$ on an FLRW background. In the algebraic approach, the quantum observables form an algebra, $\ms{A}$, which is generated by the (smeared) Klein-Gordon field $\delta\op{\phi}$ acting on a {Hilbert space} that is associated to a chosen vacuum which we will specify shortly (see, e.g., \cite{Witten:2021jzq,Hollands:2014eia} for more details).
Since the quantum field satisfies well-posed equations of motion, we may quantize the field in terms of the algebra $\Alg_{\mc{H}^{-}}$ of ``initial data'' on any Cauchy surface. The standard quantization of the inflaton perturbation and its conjugate momentum is usually done on any $t=\textrm{const.}$ Cauchy surface. It is convenient for us to instead perform the equivalent quantization on the past null boundary corresponding to the past horizon $\mc{H}^{-}=\mc{H}^{-}_{\textrm{L}}\cup \mc{H}^{-}_{\textrm{R}}$, which is a Cauchy surface for the spacetime. Choosing coordinates $(U,x^{A})$ on $\mc{H}^{-}$ where $U$ is the retarded time $U\defn t - r$ and {$x^{A}$ are arbitrary angular coordinates}, then $\delta \phi\vert_{\mc{H}^{-}}=\Phi(U,x^{A})$ corresponds to the value of the scalar field on $\mc{H}^{-}$. The initial data on $\mc{H}^{-}$ is endowed with a symplectic structure with (conserved) symplectic form, which on $\mc{H}^{-}$ takes the form
\begin{equation}
\label{eq:symp}
\Omega_{\mc{H}^{-}}(\Phi_{1},\Phi_{2}) = \frac{1}{2}\int_{\mc{H}^{-}} dUd\Omega_{2}\bigg[\Phi_{1}\partial_{U}\Phi_{2} - \Phi_{2}\partial_{U}\Phi_{1} \bigg].
\end{equation}
On $\mc{H}^{-}$ the conjugate momentum to $\Phi$ is
\begin{equation}
\label{eq:pidef}
\Pi \defn \partial_{U}\Phi .
\end{equation}
We will first consider the quantization of initial data on $\mc{H}^{-}$ which decays as $U\to \pm \infty$. This initial data corresponds to the set of modes that fully ``fall through'' the past horizon and can be entirely expressed in terms of $\Pi$. This condition excludes any zero-modes {which shall be treated separately.} The standard bulk canonical commutation relations restricted to the horizon are
\begin{align}
[\op{\Pi}(x_{1}),\op{\Pi}(x_{2})] = i \delta^{\prime}(U_{1},U_{2})\delta_{\mathbb{S}^{2}}(x^{A}_{1},x_{2}^{A}) {\op{1}}
\end{align}
{where $\op{\Pi}\in \Alg_{\mc{H}^{-}}$, $x\in \mc{H}^{-}$, and $\delta_{\bb{S}^{2}}$ is the delta function on the $2$-sphere.} The algebra $\Alg_{\mc{H}^{-}}$ admits a distinguished, translation invariant quantum state {$\Omega_{\textrm{BD}}$} (the ``Bunch-Davies state'') \cite{Chernikov:1968zm,Schomblond_1976,Bunch:1978yq,Dappiaggi:2008dk} on $\mc{H}^{-}$ corresponding to a Gaussian state with vanishing one-point function and two-point function given by\footnote{This state yields the (nearly) scale-free spectrum of density perturbations predicted by inflation.} \cite{Kay_1988} 
\begin{equation}
\label{eq:horizonstate}
\bra{\Omega_{\textrm{BD}}}\op{\Pi}(x_{1})\op{\Pi}(x_{2})\ket{\Omega_{\textrm{BD}}} = -\frac{2}{\pi}\frac{\delta_{\mathbb{S}^{2}}(x_{1}^{A},x_{2}^{A})}{(U_{1}-U_{2}-i0^{+})^{2}}.
\end{equation}
{Given the vacuum $\Omega_{\textrm{BD}}$, a Fock space, which we denote by $\Fock_{0}$, may be constructed.}
\par To obtain the initial data for the state restricted to $\ms{D}$ we choose coordinates $(u,x^{A})$ where $U=-e^{-H_{0}u}$. These coordinates cover the region $U<0$ which we denote as $\mc{H}^-_{\textrm{R}}$. If $\op{\Pi}(\omega,x^{A})$ is the Fourier transform of $\op{\Pi}(u,x^{A})$ with respect to $u$, it is straightforward to show that the $2$-point function restricted to $\mc{H}_{{R}}^{-}$ can be expressed in Fourier space as 
\begin{equation}
\label{eq:thermality}
\braket{\op{\Pi}(\omega_{1},x_{1}^{A})\op{\Pi}(\omega_{2},x_{2}^{A})}_{\Omega_{\textrm{BD}}} = \frac{2}{\pi}\frac{\omega_{1} \delta(\omega_{1},\omega_{2})\delta_{\bb{S}^{2}}(x_{1}^{A},x_{2}^{A})}{e^{\beta \omega_1}-1}
\end{equation}
where $\beta = 2\pi/H_{0}$. Thus, $\Omega_{\textrm{BD}}$ is thermal and satisfies the KMS condition with respect to translations in $u$. 

The above quantization neglected any zero-modes in the initial data which correspond to solutions of the form $\delta \phi_{0}(t)$ that satisfy 
\begin{align}
\label{eq:zm}
    \partial^2_t\delta \phi_0 + 3H  \partial_t \delta \phi_0 = -V^{\prime\prime}(\bar{\phi})\delta \phi_{0}, \quad H \defn \partial_t a/a.
\end{align}
In the asymptotic past $V^{\prime\prime}$ vanishes and the only\footnote{There also exists another linearly independent solution which exponentially diverges in the past and thus is not slowly rolling.} solution to \eqref{eq:zm} which satisfies the slow-roll conditions at early times is a mode which asymptotically approaches a constant on $\mc{H}^{-}$. Different constant modes $q\defn \delta \phi_{0}\vert_{\mc{H}^{-}}$ 
correspond to different values of the scalar during inflation and is a physical mode that must be quantized. The constant mode has a non-trivial symplectic product with any {(non-decaying)}  solution $\Phi_{p}$ whose integral 
\begin{equation}
 p \defn \int_{\mc{H}^{-}}dUd\Omega_{2}~\partial_{U}\Phi_{p} 
\end{equation}
is non-vanishing (i.e. $\Omega_{\mc{H}^{-}}({q},\Phi_{p}) \neq 0$ if $q,p\neq 0$). On phase space, the charge $p$ generates a unit shift in the zero mode $\delta \phi \to \delta \phi + \delta \phi_{0}$ where $\delta \phi_{0}\vert_{\mc{H}^{-}}=1$ \cite{Streater:1970mvf,asymp-quant,PSW-IR}. Quantization of this extended space of solutions has been extensively studied in the literature and is analogous to the ``infrared representations'' one encounters when dealing with massless fields in flat spacetimes, e.g.,  \cite{Ashtekar:1981sf,asymp-quant,Ashtekar:2018lor,Strominger:2017zoo,PSW-IR,Prabhu:2024zwl}. Here we summarize the basic results as they pertain to the quantization of the zero mode. The Fock space $\Fock_{0}$ constructed above corresponds to decaying solutions and all states in $\Fock_{0}$ have vanishing $p$ \cite{asymp-quant}. A (unitarily inequivalent) Fock space $\Fock_{p}$ of non-vanishing $p$ can be obtained by shifting the operator $\op{\Pi} \to \op{\Pi} + \partial_{U}\Phi_{p}\op{1}$ for a $\Phi_p$ with $p\neq 0$. The unique
representation that has an action of $\op{p}$ and its conjugate is the direct integral
\begin{equation}
\Hilb = \int_{\bb{R}}^{\oplus} dp~\Fock_{p} ,\quad\braket{\psi(p)|\psi(p^{\prime})} = \delta(p-p^{\prime}).
\end{equation}
{A state $\ket{\psi_{g}}\in \Hilb$ is defined by specifying $\ket{\psi(p)}\in \Fock_{p}$ together with a square-integrable wavefunction $g(p)$ on $\bb{R}$.} The operator $\op{p}$ acts as a multiplication operator which shifts the conjugate zero mode $\op{q} \defn i (d/dp)$ by a constant and satisfies $[\op{q},\op{p}]=i\op{1}$. We note that the vacuum $\ket{\Omega_{\textrm{BD}}}\in \Fock_{0}$ has a uniform distribution in the zero-mode $q$ and is therefore non-normalizable, which has been previously noted by, e.g.,~\cite{Allen_1985_scalar,Allen:1987tz,Kirsten:1993ug,Bertola:2006df,2012JCAP...11..051P}.

Finally, we canonically quantize the observer's energy 
\begin{align}
    [\op{\varepsilon} , \op{\tau}] = i \op{1},
\end{align}
which is represented on $L^2(\mathbb{R})$. 
Following \cite{Chandrasekaran:2022cip}, we impose that the observer has energy greater than its rest mass $m$. This can be straightforwardly achieved by conjugating all physical operators on $L^2(\mathbb{R})$ by $\op{P}_+$ which projects onto the subspace with $\varepsilon\geq m$.

\paragraph*{Gravitational charges and constraints.---}
Matter and geometry are intrinsically coupled through Einstein's equations, which involve constraints; therefore, the physical Hilbert space and algebra of observables are the subspaces which satisfy these constraints.

The gravitational charge, on a constant $U$ cross-section $S_U$ of $\sH^-,$ associated to any diffeomorphism $\xi^{a}$ which agrees with $u-$translation when restricted to the horizon, i.e. $\xi^{a}\vert_{\mc{H}_{\textrm{R}}^{-}}=(\partial/\partial u)^{a},$ is given by \cite{Chandrasekaran:2018aop,Ashtekar:2021wld,Hollands:2024vbe}
\begin{align}
    Q_{U,\xi} =A_U- \int_{S_U} dS~U\theta ,
\end{align}
where $\theta$ is the expansion, $A_U$ is the area of $S_U$, and $dS$ is the induced measure on $S_{U}$. We denote the $U\rightarrow \pm \infty$ limits of the charge as $Q_{\pm}$. The first-order perturbed stress energy is given by
\begin{equation}
{\delta \op{T}_{ab}= 2\dot{\bar{\phi}}t_{(a}\nabla_{b)}\delta \op{\phi} - g_{ab} (\dot{\bar{\phi}} t^{c}\nabla_{c}\delta \op{\phi} + V^{\prime}(\bar{\phi})\delta \op{\phi})} 
\end{equation}
where $\dot{\bar{\phi}}\defn \partial_{t}\bar{\phi}$. We note that the first-order energy flux $\delta \op{T}_{UU}$ through $\mc{H}^{-}$ vanishes since it is the limit of $\delta T_{ab}t^{a}t^{b}\propto g_{ab}t^{a}t^{b}$ to $\mc{H}^{-}$ where $t^{a}$ is null. Therefore, by Raychauduri's equation, the {modes that fall through the horizon} do not affect the horizon area at this order. {However, the fluctuations of the zero-mode result in order $O(G_{\textrm{N}}V^{\prime}q)$ fluctuations of the effective cosmological constant which, in turn, yield stationary fluctuations of the first-order area  \cite{2018CQGra..35o5008G}}
\begin{align}
\label{eq:firstorderconstraint2}
    \delta\op{ Q}_{+} =  \delta\op{ Q}_- = { \frac{96\pi^2  G_{\textrm{N}} V'}{(\Lambda+\Lambda_{\textrm{eff}})^2}\op{q}} \defn \frac{8\pi G_{\textrm{N}}}{H_0} \tilde{\op{q}} .
\end{align} 
At second order, the observer energy reduces the ``incoming'' area and is therefore identified with the second order perturbed charge at the quantum level \cite{Chandrasekaran:2022cip,2023arXiv230915897K}
\begin{equation}
\label{eq:secondorderconstraint1}
    \delta^2\op{ Q}_- = -\frac{8\pi G_{\textrm{N}}}{ H_{0}} \op{\varepsilon}.
\end{equation}
The energy flux through $\mc{H}^{-}$, $\delta^{2}\op{T}_{UU}\vert_{\mc{H}^{-}}= \normord{\op{\Pi}^{2}}$, is non-vanishing at second order and integrating Raychauduri's equation on $\mc{H}^{-}$ yields {a non-trivial second-order constraint }\cite{Chandrasekaran:2018aop,Hollands:2024vbe}
\begin{equation}
\label{eq:secondorderconstraint2}
    \delta^2\op{ Q}_{+} - \delta^2\op{ Q}_- = 8\pi G_{\textrm{N}}\int_{\mc{H}^-}dUd\Omega_2U \delta^2 \op{T}_{UU}\defn  - \frac{8\pi G_{\textrm{N}}}{H_0}\op{F}_\xi
\end{equation}
where $\op{F}_{\xi}$ generates infinitesimal translations in $u$. 
The zero-mode does not contribute at this order because $V^{\prime \prime}(\bar{\phi})\to 0$ as $t\to -\infty$. 

\paragraph*{Crossed product algebra and generalized entropy.---}
We are now in a position to consider the algebra of dressed observables associated to the observer, i.e.~those {accessible to the observer and} commuting with the gravitational constraints.
The observer has access to the quantum fields and the charges located in $\mathscr{D}$. 
The first order constraint, \eqref{eq:firstorderconstraint2}, tells us that $\delta\op{ Q}_-$ is identified with the zero-mode (and $\delta\op{ Q}_+$) while the second order constraint \eqref{eq:secondorderconstraint1} identifies $\delta^2\op{ Q}_-$ and $\op{\varepsilon}$. We are not yet done because $\delta^2\op{Q}_+$ is spacelike separated to $\ms{D}$, however $\delta \op{\phi}$ does not commute with  $\delta^2\op{Q}_+ $ due to the second order constraint \eqref{eq:secondorderconstraint2}. The subalgebra of operators in $\ms{D}$ that satisfy the constraints is
\begin{align}
\label{eq:dress}
    \Alg_{\textrm{dress.}} \defn \op{P}_+\{e^{-i\op{F}_{\xi}\op{\tau}}\delta \op{\phi}e^{i\op{F}_{\xi}\op{\tau}},  \tilde{\op{q}} ,\op{\varepsilon}\}''\op{P}_+
\end{align}
where $'$ is the commutant i.e.~the set of bounded operators on the Hilbert space that commute with the algebra. 
To make the algebra a so-called von Neumann algebra, we took the commutant twice.
The reason to use von Neumann algebras is that they enable us to compute
von Neumann entropy. We will only introduce the essential elements of von Neumann algebra theory (see, e.g., section 3.2 of \cite{2023arXiv230915897K} or \cite{2018arXiv180304993W, 2023arXiv230201958S} for more details).

This algebra has an interesting structure. The first thing to note is that it has a center; $\tilde{\op{q}}$ commutes with all other elements. The second important feature is that the algebra has the structure of a crossed product, $\Alg_{\textrm{dress.}} = (\Alg_{\mc{H}_{\textrm{R}}^{-}} \rtimes \Alg_{\varepsilon})\otimes \Alg_{\tilde{q}}$. Since there exists a state on the QFT operators that is thermal with respect to time translations by $\op{F}_{\xi}$, \eqref{eq:thermality}, a theorem of Takesaki \cite{takesaki1973duality} tells us that the algebra is Type II$_{\infty}$. The reason this is important is that Type II algebras --- unlike the Type III algebras that are arise in non-gravitational QFT --- admit traces. Namely,
\begin{align}
\begin{aligned}
\label{eq:tr}
     \Tr(\hat{a}) = \bra{\Omega_{\textrm{BD}},0_\tau } e^{\frac{2\pi}{ H_{0}}(\tilde{\op{q}} -\op{\varepsilon})} \hat{a}\ket{\Omega_{\textrm{BD}},0_\tau } 
\end{aligned}
\end{align}
is a trace for $\hat{a} \in \Alg_{\textrm{dress.}}$ where $\ket{0_{\tau}}$ is the (improper) eigenstate of $\op{\tau}$ with zero eigenvalue \cite{2022JHEP...10..008W}.
That the algebra is Type II$_\infty$, rather than Type II$_1$ means that the trace of the identity is infinite. Due to the center, the trace is only unique up to functions of $\tilde{\op{q}}$.
As we will see, the specific choice of \eqref{eq:tr} will imply that the von Neumann entropy of any semiclassical state is the generalized entropy. We expect that working to higher orders in perturbation theory will render the trace unique and equivalent to \eqref{eq:tr}. 

With a trace, we can define density matrices and von Neumann entropies for states via the defining relations
\begin{align}
    \Tr (\rho_\Psi \hat a) = \bra{\Psi} \hat{a}\ket{\Psi}, \quad S_{\textrm{vN}}(\rho_\Psi) \defn -\Tr \rho_\Psi \log \rho_\Psi.
\end{align}
Taking $\ket{\Psi} =\int d\varepsilon f(\varepsilon)\ket{\psi_g}\ket{\varepsilon}$, 
the density matrix is given by \cite{Chandrasekaran:2022cip,Jensen:2023yxy}
\begin{align}
    \rho_{\Psi} &= f(\op{\varepsilon})e^{-i\op{\tau}\op{F}_{\xi}}e^{-\frac{2\pi}{ H_{0}}(\tilde{\op{q}}-\op{\varepsilon})}\Delta_{\Omega_{\textrm{BD}}|\psi_g}e^{i\op{\tau}\op{F}_{\xi}}f(\op{\varepsilon}),
\end{align}
where for any two states $\omega, \psi \in \mathscr{H}$, $\Delta_{\s|\psi} \defn S_{\s|\psi}^{\dagger}S_{\s|\psi}$ is the relative modular operator, which is defined in terms of the relative Tomita operator which satisfies \cite{takesaki2002theory}
\begin{align}
    S_{\s|\psi} a \ket{\psi} = a^{\dagger} \ket{\s}, \quad \forall a \in \Alg.
\end{align}
Assuming $f$ is slowly varying in its argument, the von Neumann entropy is then \cite{Chandrasekaran:2022cip,2023arXiv231207646K}
\begin{align}
\label{eq:svnfinal}
    S_{\textrm{vN}}(\rho_{\Psi}) = \frac{2\pi}{ H_{0}}\langle\tilde{\op{q}}-\op{\varepsilon}\rangle_{\Psi}-S_{\textrm{rel.}}(\psi_g|\Omega_{\textrm{BD}})+S_f ,
\end{align}
where the first term is the change in entropy due to the $O(G_{\textrm{N}})$ 
change of area of the horizon, the second term is the relative entropy, $S_{\textrm{rel.}}\defn -\bra{\s} \log \Delta_{\s|\psi} \ket{\s} $, of the quantum fields and the final term is the Shannon entropy of the probability distribution $|f|^2$. All terms are finite but there is no upper bound on the entropy because $\tilde{\op{q}}$ is $\mathbb{R}$ valued, a manifestation of the Type II$_{\infty}$ nature of the algebra. One may heuristically rewrite the relative entropy term such that the right-hand side of \eqref{eq:svnfinal} takes the more familiar form of the generalized entropy associated to the comoving observer \eqref{eq:sgen} up to a state-independent constant \cite{2012PhRvD..85j4049W,2022arXiv220910454C,2023arXiv230915897K}. Interestingly, this entropy has been recently given a (relative) state-counting interpretation \cite{2024arXiv240416098A}.

\paragraph*{Discussion.---}In this Letter, we have constructed the algebra of observables for an observer in a universe like our own, assuming an initial period of inflation driven by a scalar field. For the sake of clarity, we described a simplified model consisting only of the metric, inflaton, and observer, though all of our conclusions fully generalize to interacting fields, such as the full field content of the standard model (see SM for details). We found that due to the presence of the cosmological horizon, the observer has an 
entropy equaling the generalized entropy of their causal diamond $\ms{D}$, which is sensitive to the area of the horizon and initial conditions of the field content at the beginning of inflation.

Our analysis of FLRW spacetimes is suggestive that for all spacetimes with horizons, the algebra associated to an observer is Type II. In the exceptional case of empty de Sitter without a dynamical cosmological constant, the algebra is Type II$_1$ \cite{Chandrasekaran:2022cip} and there exists a state of maximum entropy, but in all other cases, we believe it is reasonable
to hypothesize that the algebra is Type II$_{\infty} $ and the entropy is unbounded, as discussed in \cite{2023arXiv230803663W} and demonstrated in \cite{2023arXiv230915897K} and the present Letter. 

It is interesting to speculate if and how the story is modified when working to higher orders in gravitational perturbation theory and eventually non-perturbative physics. {In perturbation theory,} we expect that the resulting gauge invariant algebras will remain Type II$_{\infty}$ but will not have a center, rendering the trace unique. We hope this will lead to a better understanding of cosmological observables.

\paragraph*{Note Added.---}The submission of this paper to arXiv has been coordinated with the submission of \cite{penington_chen}, which considers the algebra of observables in a spacetime where \eqref{eq:V(phi)} holds for \textit{all} times, such that inflation is eternal and the background is exact de Sitter. In this special case, one can construct dressed observables by using the full rolling field as a ``clock.''

\acknowledgments

We thank Robert Brandenberger, Chang-Han Chen, Juan Maldacena, Vladimir Narovlansky, Geoff Penington, Kartik Prabhu, Jonathan Sorce, and Aron Wall for discussions and comments. We are particularly grateful to Edward Witten for discussions during the early stages of this work.
JKF is supported by the Marvin L.~Goldberger Member Fund at the Institute for Advanced Study and the National Science Foundation under Grant No. PHY-2207584. S.L.~acknowledges the support of NSF grant No. PHY-2209997. G.S.~and S.L.~are supported by the Princeton Gravity Initiative at Princeton University.

\bibliographystyle{apsrev4-2}
\bibliography{main}

\appendix

\onecolumngrid

\section{Relation to existing literature}

Here, we briefly comment on papers with some relation to this work (see also \cite{2023EPJC...83.1003S,2022arXiv220706704G,2023arXiv230214747G}).

\begin{itemize}
    \item \cite{Chandrasekaran:2022cip}: In this work, the algebra of gravitationally-dressed observables is constructed for pure de Sitter space. The background is sourced by a non-fluctuating cosmological constant, so there is no zero-mode and the algebra is Type II$_1$. If one were to quantize the zero-mode of a scalar field in this background, the energy due to fluctuations of this mode would be O$(V^{\prime\prime}q^{2})$. For any stable field theory, $V^{\prime\prime}>0,$ so the energy is positive and the algebra will not be deformed.
    \item \cite{Jensen:2023yxy}: In this work, the algebra of gravitationally-dressed observables is constructed for general spacetime regions under the following set of assumptions
    \begin{itemize}
    \item[(A1)] There exists a quantum field theory algebra of observables, $\Alg$, associated to a causally complete region, $\mathscr{D}$, that is type III$_1$.
    \item[(A2)] There is an auxiliary type I$_{\infty}$ algebra of an observer associated to $\mathscr{D}$.
    \item[(A3)] The physical algebra arises from imposing the gravitational constraint that operators commute with $F_{\xi} + q$.
    \item[(A4)] The flow generated by $F_{\xi}$ coincides with the modular flow for some state on the QFT algebra
    \item[(A5)] The gravitational constraints hold on a Cauchy surface
    \item[(A6)] The energies of observers are bounded from below.
    \item[(A7)] The region $\ms{D}$ can be defined in a diffeomorphism invariant way. 
    \item[(A8)] There are no other algebras of observables associated to the region 
\end{itemize}
While (A7) and (A8) were not explicitly written in \cite{Jensen:2023yxy}, it was implicit. From these assumptions, the authors demonstrated that for bounded subregions, the algebra is Type II$_1$. While assumptions (A1) - (A7) hold in the cosmological spacetimes that we analyze, assumption (A8) does not. The additional algebra of observables is, of course, the zero-mode, which leads to a Type II$_{\infty}$ algebra. (A8) similarly did not hold in \cite{2023arXiv230915897K} for the algebra of Schwarzschild-de Sitter. It is reasonable to expect that (A8) always fails when the background geometry is a member of a continuous family of solutions to Einstein's equations. As emphasized in \cite{2023arXiv230803663W}, we expect that quantum fluctuations in this continuous family will generally yield a Type II$_{\infty}$ algebra.
    \item \cite{2023arXiv230803663W} In this work, an algebra of dressed observables was constructed for an observer without the specification of a spacetime background. It was suggested that for closed universes where the Hartle-Hawking state is a normalizable state, the algebra is Type II$_1$ and for spacetimes where it is non-normalizable, it is Type II$_{\infty}$. Motivation for this came from considerations of Schwarzschild de Sitter and backgrounds with leading order backreaction from matter. The conclusions are consistent with the current paper.
\end{itemize}

\section{Quantum gravitational perturbations and general matter}
\label{appendixA}
The observations we have made of our universe appear to be well-described by $\Lambda$CDM together with the standard model. However in the main text, for simplicity, we considered a minimal model of a cosmological spacetime with metric $g_{ab}$, an observer $\gamma$, and inflaton field $\phi$. Furthermore, we explicitly considered the effects of $O(\hbar)$ fluctuations of the inflaton field and their backreaction on the spacetime geometry. This analysis ignored all other effects in the early universe due to the standard model as well as any other cosmologically relevant fields.  In this section we will illustrate that, under minimal assumptions, the inclusion of any other fields does not fundamentally change the conclusion of this Letter. In particular, we continue to assume that the other fields only have $O(G_{\textrm{N}}^0)$ stress-energy in the asymptotic past (i.e. the only degree of freedom with large backreaction is the inflaton prior to reheating).

\subsubsection{Free Fields}
Prior to analyzing the effects of general matter, we note that the analysis in the main text considered the leading order backreaction effects of the inflaton and neglected any contributions due to gravitational fluctuations which are present even in our minimal model. In particular, we must also consider small $O(\hbar^{1/2}G_{\textrm{N}}^{1/2})$ fluctuations in the metric 
\begin{equation}
g_{ab} = \bar{g}_{ab} + \delta g_{ab}
\end{equation}
where $\bar{g}_{ab}$ is the $O(\hbar^{0}G_{\textrm{N}}^{0})$ ``background metric''. These metric fluctuations satisfy the linearized Einstein equations 
\begin{align}
    -\frac{1}{2}\Box_{\bar{g}}\delta g_{ab} -\frac{1}{2}\nabla_{a}\nabla_{b}\delta g^{c}{}_{c}+\frac{1}{2}\nabla_{c}\nabla_{a}\delta g^{c}{}_{b}+\frac{1}{2}\nabla_{c}\nabla_{b}\delta g^{c}{}_{a}-\Lambda \delta g_{ab}=0 
\end{align}
where indices are raised and lowered with the background metric $\bar{g}_{ab}$. It is straightforward to quantize the perturbed metric which we denote as $\delta \op{g}_{ab}$ (see, e.g., Appendix C of \cite{2023arXiv230915897K} for more details). The initial data for the graviton perturbations on $\mc{H}^{-}$ correspond to the algebra $\Alg^{\textrm{GR}}$ generated by the perturbed shear on the horizon $\delta \op{\sigma}_{AB}$ where capital indices refer tensor indices on the $2$-sphere cross-sections of the horizon. The commutation relations are given by 
\begin{align}
    [\delta \op{\sigma}_{AB}(x_1) ,\delta \op{\sigma}_{CD}(x_2) ] =  \frac{i}{2} \delta^{\prime}(U_{1},U_{2})\delta_{\mathbb{S}^{2}}(x^{A}_{1},x_{2}^{A})\bigg(q_{A(C}q_{D)B}-\frac{1}{2}q_{AB}q_{CD}\bigg)\op{1}.
\end{align}
where $q_{AB}$ is the round metric on the sphere.
The initial data admits a distinguished, translation invariant, Gaussian quantum state $\Omega_{\textrm{BD}}^{\textrm{GR}}$ which corresponds to the analog of the Bunch-Davies state for gravitational perturbations \cite{Allen:1986_graviton}. The $2$-point function in this state on $\mc{H}^{-}$ is  
\begin{equation}
\label{eq:horizonstategrav}
\braket{\Omega_{\textrm{BD}}^{\textrm{GR}}|\delta \op{\sigma}_{AB}(x_{1})\delta \op{\sigma}_{CD}(x_{2})|\Omega_{\textrm{BD}}^{\textrm{GR}}}=\frac{1}{\pi}\frac{(q_{A(C}q_{D)B}-\frac{1}{2}q_{AB}q_{CD})\delta_{\bb{S}^{2}}(x_{1}^{A},x_{2}^{A})}{(U_{1}-U_{2}-i0^{+})^{2}}.
\end{equation}
which is analogous to \eqref{eq:secondorderconstraint2}. By the same arguments as in the main text, the restriction of $\Omega_{\textrm{BD}}^{\textrm{GR}}$ to $\mc{H}^{-}_{\textrm{R}}$ is a thermal state with respect to $\xi^{a}\vert_{\mc{H}^{-}}=(\partial/\partial u)^{a}$ with inverse temperature $2\pi/H_{0}$. Given this vacuum, one can construct the corresponding Fock space of gravitons $\Fock^{\textrm{GR}}$ so that the full Hilbert space $\Hilb$ in the main text is only changed by  $\Hilb \otimes \Fock^{\textrm{GR}}$. Furthermore, any ``zero mode'' of the graviton is pure gauge \cite{Higuchi:2000ye,Higuchi:2001uv,Hawking:2000ee,Kouris:2001hz} and so the gravitational constraints considered in the main text now get an additional contribution to \eqref{eq:secondorderconstraint2} due to the energy flux $\op{F}_{\xi}^{\textrm{GR}}$ of gravitons through the horizon 
\begin{equation}
\op{F}_{\xi}^{\textrm{GR}} = -\frac{H_0}{4\pi G_{\textrm{N}}}\int_{\mc{H}^{-}}dUd\Omega_{2}~U:\delta \op{\sigma}_{AB}\delta \op{\sigma}^{AB}:.
\end{equation}
Imposing the gravitational constraints on the algebra imply that the graviton perturbations are gravitationally dressed by conjugating $\delta \op{g}_{ab}$ by $e^{i\op{F}^{\textrm{GR}}_{\xi}\op{p}}$ and the full dressed algebra becomes $([\Alg_{\sH^{-}_{\rm R}} \otimes \Alg^{\textrm{GR}}_{\sH^{-}_{\rm R}}] \rtimes \Alg_{\varepsilon})\otimes \Alg_{\tilde{q}}$. By the thermality of the vacuum $\Omega_{\textrm{BD}}^{\textrm{GR}}$, this algebra is again a Type II$_{\infty}$ algebra and the generalized entropy of any semiclassical state is given by \eqref{eq:sgen}. 

We now consider the effects of additional matter fields. While we largely focus on standard model fields, the present discussion applies, in principle, to any degrees of freedom present in the early universe. As emphasized in the main text, the algebra of observables in $\ms{D}$ can be straightforwardly obtained by considering the algebra of observables on the past horizon $\mc{H}^{-}$ which is essentially\footnote{We note that $\mc{H}^{-}$ does not admit the strict requirements of a Cauchy surface since there exist timelike curves which hit past timelike infinity and never pass through $\mc{H}^{-}$. However, $\mc{H}^{-}$ is a good initial data surface for wave propagation since all modes eventually fall through the past horizon.}  a Cauchy surface for $\ms{D}$. We will first, for simplicity, neglect their interactions and simply consider the contribution in the free field approximation. We will then illustrate that the conclusions remain valid in the interacting theory. 

At the earliest moments of inflation, the temperature of the universe was far greater than the quark-hadron phase transition (i.e., $T\gg 150 \textrm{MeV}$) and therefore the quarks and gluons are deconfined at these temperatures. The commutation relations of any Yang-Mills fields $\op{E}_{A}^{i}$ with any (semi-simple) gauge group $G$ on $\mc{H}^{i}$ are given by \cite{PSW-IR}
\begin{equation}
[\op{E}^{i}_{A}(x_{1}),\op{E}^{j}_{B}(x_{2})] = -\frac{i\pi}{2} \delta^{\prime}(U_{1},U_{2})\delta_{\mathbb{S}^{2}}(x^{A}_{1},x_{2}^{A})k^{ij}q_{AB}\op{1}.
\end{equation}
where $i,j$ denote Lie-algebra indices, $k_{ij}=-c^{l}{}_{ik}c^{k}{}_{jl}$ is the Killing-Cartan metric and $c^{i}{}_{ij}$ are the structure constants of the Lie algebra $\mf{g}$. The $2$-point function for the ``Bunch-Davies'' vacuum is
\begin{equation}
\label{eq:horizonstateYM}
\braket{\Omega_{\textrm{BD}}^{\textrm{YM}}|\op{E}^{i}_{A}(x_{1})\op{E}^{j}_{B}(x_{2})|\Omega_{\textrm{BD}}^{\textrm{YM}}}=-\frac{k^{ij}q_{AB}\delta_{\bb{S}^{2}}(x_{1}^{A},x_{2}^{A})}{(U_{1}-U_{2}-i0^{+})^{2}}.
\end{equation}
For gluon fields the relevant gauge group is $\textrm{SU}(3)$ whereas for electromagnetic fields the gauge group is $\textrm{U}(1)$. The quantization of any free spinor fields (e.g. quarks or electrons) can be treated similarly (see section 4.1 of \cite{2012PhRvD..85j4049W} for details). 
As in the case of gravitons, the corresponding Fock space is straightforwardly included in the Hilbert space. These fields contribute to the second order gravitational constraints via their energy fluxes given by $\op{F}_{\xi}^{\textrm{EM}}$, $\op{F}_{\xi}^{\textrm{YM}}$ as well as the spinor fluxes through $\mc{H}^{-}$. These fluxes generate time-translations on $\mc{H}^{-}$ on their respective Hilbert spaces and the vacua are thermal with respect to these translations. These fluxes contribute to the gravitational constraints \eqref{eq:secondorderconstraint2} and imply that each field is dressed to the observer in an analogous manner as described above. 

\subsubsection{Interacting Fields}

Up until now, we have considered the effects of any free quantum fields in a cosmological spacetime. The interactions of the fields initially present in the early universe are of course relevant to describe the universe we observe such as, reheating as the inflaton energy is converted into radiation, the hadronization of quarks and gluons, and structure formation. However, the above arguments must be slightly modified if one wishes to consider any interacting quantum field theory. The main technical ingredients for our arguments in the main text were the gravitational constraint \eqref{eq:secondorderconstraint2} and the existence of a translation invariant state $\Omega_{\textrm{BD}}$. The existence of such a state directly implies that when restricted to the algebra $\Alg_{\mc{H}_{\textrm{R}}^{-}}$ of initial data on $\mc{H}_{\textrm{R}}^{-}$, such a state is thermal with respect to the ``boost-like'' vector field $\xi$ on $\mc{H}^{-}$ considered in the main text. By Takesaki's theorem \cite{takesaki1973duality}, these two ingredients directly imply that the ``dressed algebra'' is Type II. The main technical hurdle in directly applying these arguments to any interacting theory is that one cannot define an algebra of observables on $\mc{H}^{-}$ simply by the restriction\footnote{As noted in \cite{2012PhRvD..85j4049W}, any superrenormalizable theory can be directly restricted to a null surface.} of fields to the null surface since operators restricted in this way will generally have infinite fluctuations \cite{2012PhRvD..85j4049W}. This issue was recently addressed in our present context by Faulkner and Speranza \cite{Faulkner:2024gst} using arguments by Casini, Teste and Torroba \cite{Casini:2017roe} as well as Witten \cite{Witten_lecture}. We now sketch the basic arguments of these constructions (see, e.g., section 3.1. \cite{Faulkner:2024gst} for details). 

\begin{figure}
    \centering
    \includegraphics[]{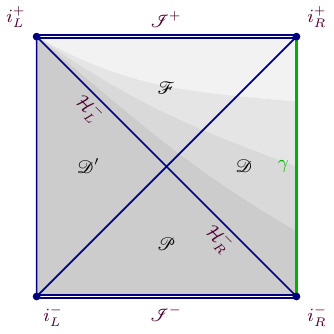}
    \caption{An example of a globally hyperbolic extension $\bar{M}$ of the spacetime $M$ considered in this paper such that $\bar{M}$ is geodesically complete. This can be achieved by gluing the de Sitter region $\ms{D}' \cup \ms{P}$ to $M=\mc{F}\cup \ms{D}$.
    }
    \label{fig:penrose_extend_tikz}
\end{figure}

To present the argument, it will be technically convenient to first extend the spacetime in the following way. We note that the spacetimes $(M,g)$ considered in this Letter are not past complete --- past-directed null geodesics reach $\mc{H}^{-}$ in finite affine time. Since, the spacetime is asymptotically de Sitter approaching $\mc{H}^{-}$, we can (non-uniquely) extend the spacetime to a smooth, complete spacetime $\bar{M}$ which remains globally hyperbolic. While the particular choice of extension will not be relevant to our arguments one could, for example, smoothly attach the ``bottom'' Poincaré patch of de Sitter spacetime at $\mc{H}^{-}$, $\ms{D}'\cup \ms{P}$ to $M$ (see Figure \ref{fig:penrose_extend_tikz}). The surface $\mc{H}^{-}$ is a Cauchy surface for the completed spacetime and so all quantum states initially defined in $M$ are regular 
on $\mc{H}^{-}$ and so the state on $\bar{M}$ is uniquely defined by evolution. 

In order to define an algebra of operators associated to a null surface, we first analyze integrals of the (renormalized) energy flux through $\mc{H}^{-}$. The (second-order) ANEC and boost operators are
\begin{equation}
\op{F}(x^{A}) \defn \int_{-\infty}^{\infty}dU~\delta^{2}\op{T}_{UU}(U,x^A) \quad \textrm{ and }\quad \op{F}_{\xi}\defn \int_{-\infty}^{\infty}dUd\Omega_{2}~U\delta^{2}\op{T}_{UU}(U,x^A) .
\end{equation}
Even though these operators are localized to the horizon, they are well-defined elements of the algebra that translate or boost operators localized to $\mc{H}^{-}$. If the quantum field theory satisfies the average null energy condition, then $\op{F}(x^A)$ is a positive operator and the ``Bunch-Davies vacuum'' $\ket{\Omega_{\textrm{BD}}}$ is defined as the ground state which is annihilated by this operator for any $x^A$. $\op{F}(x^A)$ and $\op{F}_{\xi}$ commute and $\ket{\Omega_{\textrm{BD}}}$ will also be annihilated by $\op{F}_{\xi}$. The ANEC has been proven for interacting quantum field theories in flat spacetime \cite{2016JHEP...09..038F,2017JHEP...07..066H} and, in curved spacetime, is expected to hold on any Killing horizon.

While a general algebra element cannot be restricted to $\mc{H}^-$, one can associate an algebra of operators to the horizon by noting that matrix elements of all horizon-localized operators\footnote{These are more precisely referred to as sesquilinear forms.} 
are well-defined. If $\ms{S}_{\mc{H}_{\textrm{R}}^{-}}$ is the set of such operators, then its commutant $\ms{S}^{\prime}_{\mc{H}_{\textrm{R}}^{-}}$ is the algebra of (bounded) operators localized in the causal complement of $\ms{D}$ which we denote as $\ms{D}^{\prime}$. Taking an additional commutant yields a (von Neumann) algebra $\ms{A}_{\ms{D}}$ of operators in $\ms{D}$. Thus, the operators $\ms{S}_{\mc{H}_{\textrm{R}}^{-}}$ localized on the horizon, generate the algebra in $\ms{D}$. Furthermore, since $\op{F}_{\xi}$ generates boosts, it is straightforward to show that it preserves $\ms{A}_{\ms{D}}$. Using this fact, 
one can show that $\Omega_{\textrm{BD}}$ is thermal\footnote{More precisely, the modular Hamiltonian of $\ket{\Omega_{\textrm{BD}}}$ on $\ms{A}_{\ms{D}}$ coincides with $\op{F}_{\xi}$ on $\mc{H}^{-}$.} on the elements of $\ms{S}_{\mc{H}_{\textrm{R}}^{-}}$ \cite{Borchers:1991xk,Borchers:2000pv,Faulkner:2024gst}. 

Thus, we have succeeded in obtaining a state $\ket{\Omega_{\textrm{BD}}}$ which is translation invariant on $\mc{H}^{-}$ and thermal with respect to boosts on $\mc{H}_{\textrm{R}}^{-}$. The existence of $\op{F}_{\xi}$ implies that the gravitational constraints \eqref{eq:secondorderconstraint2} are well-defined and imposing these constraints yields an algebra of dressed observables which is again Type II. Thus, assuming that the ANEC is satisfied, the addition of any other matter fields does not change the conclusions of this paper. Due to the zero-mode of the inflaton, the algebra will remain Type II$_\infty$ and the von Neumann entropy of any semiclassical state is the generalized entropy \eqref{eq:sgen}.

\end{document}